\documentclass[aps,12pt]{revtex4-2}

\usepackage{graphicx}
\usepackage{dcolumn}
\usepackage{amsmath}
\usepackage{amssymb}
\usepackage{amsbsy}
\usepackage{float}
\usepackage{url}
\usepackage[bookmarksopen]{hyperref}
\usepackage{caption}
\usepackage{physics}
\usepackage{float}
\usepackage{natbib}

\begin{document}

\title{Quantum Grid Path Planning Using Parallel QAOA Circuits Based on Minimum Energy Principle}

\author{Liu Jun}
\email{liujun@hufe.edu.cn}
\affiliation{School of Economics, Hunan University of Finance and Economics, Changsha 410205, China }

\begin{abstract}
	\textbf{Abstract} To overcome the bottleneck of classical path planning schemes in solving NP problems and address the predicament faced by current mainstream quantum path planning frameworks in the Noisy Intermediate-Scale Quantum (NISQ) era, this study attempts to construct a quantum path planning solution based on parallel Quantum Approximate Optimization Algorithm (QAOA) architecture. Specifically, the grid path planning problem is mapped to the problem of finding the minimum quantum energy state. Two parallel QAOA circuits are built to simultaneously execute two solution processes, namely connectivity energy calculation and path energy calculation. A classical algorithm is employed to filter out unreasonable solutions of connectivity energy, and finally, the approximate optimal solution to the path planning problem is obtained by merging the calculation results of the two parallel circuits. The research findings indicate that by setting appropriate filter parameters, quantum states corresponding to position points with extremely low occurrence probabilities can be effectively filtered out, thereby increasing the probability of obtaining the target quantum state. Even when the circuit layer number $p$ is only 1, the theoretical solution of the optimal path coding combination can still be found by leveraging the critical role of the filter. Compared with serial circuits, parallel circuits exhibit a significant advantage, as they can find the optimal feasible path coding combination with the highest probability.
    
\end{abstract}

	\maketitle
	\section{Introduction}
	The grid path planning problem can essentially be viewed as an optimal path problem in graph theory. Each node in the grid represents a potential location that an agent can reach, with adjacent nodes connected by edges; the path cost of an edge is proportional to the obstacles the agent encounters while moving between the corresponding nodes. Dijkstra (1959) was the first to develop a classical optimal path algorithm for grid maps. Based on a greedy strategy, this algorithm prioritizes expanding the path with the minimum current cost, ensuring the discovery of a globally optimal solution in the map\cite{Dijkstra1959}. Building on Dijkstra’s algorithm, Hart et al. (1968) proposed the A* algorithm, marking the formal application of heuristic search in grid path planning and establishing it as a core method in subsequent research on this topic\cite{Hart1968}. Stentz (1994) further designed the D* algorithm, which enables efficient optimal path planning in unknown and dynamic environments\cite{Stentz1994}.
	
	In recent years, relevant researches have achieved further progress. Luo et al. (2020) extended the traditional Dijkstra algorithm by equivalently converting surface triangular grids into 2D planar triangles to solve for optimal paths; simulation results confirmed that this approach significantly improves the accuracy of surface-optimized paths\cite{Luo2020}. Dang (2020) attempted to design several improved suboptimal path planning techniques for 2D grid maps. These techniques yield paths whose lengths are asymptotically close to the optimal solution, while their runtime is significantly reduced compared to the traditional A* algorithm\cite{Dang2020}. Addressing the multi-robot path planning problem based on grid maps, Han et al. (2020) proposed the DDM algorithm. Through two heuristic designs—multiple paths and a database for solving optimal subproblems—this algorithm enables rapid solution of large-scale labeled path planning tasks with near-suboptimal performance\cite{Han2020}. Zhou (2021) developed a computational process to convert 3D maps into 2D grid maps in real time, followed by the application of the A* algorithm to find optimal paths\cite{Zhou2021}.
	
	With the rise of artificial intelligence and machine learning, neural networks and reinforcement learning methods have been widely adopted in solving grid map path planning problems. Babu et al. (2016) were among the early researchers to use Q-learning for autonomous robot navigation in unknown environments, demonstrating the feasibility of applying reinforcement learning to optimal path planning\cite{Babu2016}. Panov et al. (2018) became the first to apply deep reinforcement learning to grid map path planning, verifying its practical utility in fields such as robot navigation and video games\cite{Panov2018}. Gao et al. (2020) designed an incremental training framework that combines deep reinforcement learning with traditional probabilistic roadmaps, effectively enhancing the path planning capability of indoor mobile robots in 2D grid maps\cite{Gao2020a}. Li et al. (2022) proposed an improved deep Q-network (DQN) path planning algorithm; by introducing a prioritized experience replay mechanism and a dual-network structure, this algorithm improved both the learning efficiency in high-dimensional state spaces and the accuracy of grid path planning\cite{Li2022}. Malczyk (2025) leveraged a semantic-aware deep reinforcement learning framework to enhance the generalization ability of optimal path planning for complex inspection tasks, exhibiting strong robustness particularly under conditions involving dynamic obstacles\cite{Malczyk2025}.
	
	However, the grid path planning problem is inherently an NP problem, and the efficiency of existing algorithms decreases significantly on large-scale grid maps. Even theoretically, when the scale of the grid map reaches a certain level, classical algorithms will be unable to compute and solve its optimal path planning problem. In recent years, the continuous development of quantum computing and quantum algorithms has provided new ideas for breaking this bottleneck. The core of quantum optimal path planning lies in mapping the optimal path solving problem into a quantum computing process, and accelerating the solving process by giving full play to the advantages of quantum computing.
	
	In terms of relevant technical foundations, D-Wave (2011) took the lead in implementing the optimization of the Ising model on a quantum annealer. In the experiment, by programming the coupling strength between qubits, the solution of NP problems in classical computing was simulated, demonstrating the potential to outperform classical computing in combinatorial optimization problems\cite{Johnson2011}. Tencent Quantum Laboratory (2022) proposed an enhancement method combining Monte Carlo Tree Search (MCTS) with a neural network to optimize the scheduling strategy of quantum annealing and improve its performance, which showed excellent performance in solving various combinatorial optimization tasks\cite{Chen2022}. These studies provide sufficient technical support for solving quantum path planning problems. In terms of early exploratory research, Santha (2008) systematically analyzed the application of quantum walks in unordered database search, derived the quadratic acceleration condition based on this, and designed a quantum algorithm framework for improving node access efficiency in path planning problems\cite{Santha2008}. Weber et al. (2014) verified the optimal path prediction theory of quantum system state transition through experiments, proving that the quantum trajectory is consistent with the theoretical prediction, which provides a physical implementation basis for quantum path planning\cite{Weber2014}.
	
	In recent years, research results in the field of quantum path planning have been emerging continuously. Gao et al. (2020) integrated quantum computing with optimization algorithms, and used the characteristics of quantum superposition and entanglement to improve the global optimization capability of path search. In their study, quantum encoding was adopted to represent path information, and an adaptive quantum rotation gate adjustment strategy was designed to improve convergence efficiency\cite{Gao2020b}. Chella (2022) designed a quantum path planner based on Grover's algorithm to realize robot motion planning. It was found in a $4\times4$ map that when the average branching factor of the search tree is high, the efficiency of the quantum algorithm is superior to that of classical methods\cite{Chella2022}. Azad et al. (2023) applied the QAOA algorithm to the vehicle path planning problem and encoded path constraints into a Hamiltonian. Experiments have proved that the improved Hamiltonian can significantly improve the solution accuracy and reduce the probability of local optimal solutions\cite{Azad2023}. Tran et al. (2023) designed a quantum-classical hybrid architecture for solving dynamic path planning problems, where the quantum layer performs QUBO optimization and the classical layer processes environmental feedback. In the UAV obstacle avoidance experiment, it was found that the path re-planning delay was significantly reduced, and the energy consumption was significantly decreased\cite{Tran2023}. Chella et al. (2023) combined quantum computing principles with swarm intelligence algorithms to improve the search efficiency and global optimization capability of the algorithm in complex dynamic environments. Compared with traditional swarm intelligence algorithms, it showed better performance in terms of convergence speed and solution quality\cite{Chella2023}. Li (2025) focused on the path-finding problem in unweighted undirected graphs, constructed a graph structure based on a welded tree, and proposed an efficient quantum algorithm to find the optimal path between two points. The study proves that classical algorithms are difficult to solve with high probability in sub-exponential time, while quantum algorithms can achieve exponential acceleration\cite{Li2025}.
	
	A further review of the literature reveals that, regarding the technical pathways for solving quantum path planning problems, researchers widely adopt schemes based on quantum ground states. The fundamental idea is to map the path planning problem to the problem of finding the ground state energy of the Ising model. Position nodes in the grid are encoded as corresponding qubits, while the connectivity patterns between nodes and the path costs are characterized by constructing a Hamiltonian that incorporates constraint conditions. In terms of specific implementation, the current mainstream approach involves sequentially encoding the main component of the Hamiltonian and the constraint conditions into a single quantum circuit for serial operation. However, under the existing NISQ conditions, the depth and complexity of the quantum circuit have a direct impact on computational accuracy. When the number of position nodes is large, the circuit depth and complexity increase accordingly, which in turn impairs computational precision.
	
	To address the aforementioned limitations of mainstream algorithms, this study proposes that two parallel QAOA circuits can be constructed to simultaneously perform two solution processes: connectivity energy calculation and path energy calculation. Subsequently, classical algorithms are used to merge the computational results of the two processes, ultimately obtaining an approximately optimal solution for the path planning problem. The subsequent research logic of this study is as follows: First, the basic framework for problem solving is defined; second, the connectivity conditions of the grid map are characterized, and the basic steps of the overall solution process are further outlined; third, a technical scheme for problem solving based on the parallel QAOA architecture is designed, and on this basis, computational experiments are conducted to solve the optimal path for a \(2 \times 3\) grid map.

	\section{Basic Framework for Problem Solving}
	
\begin{figure}[h]
		\centering
		\includegraphics[width=1\linewidth]{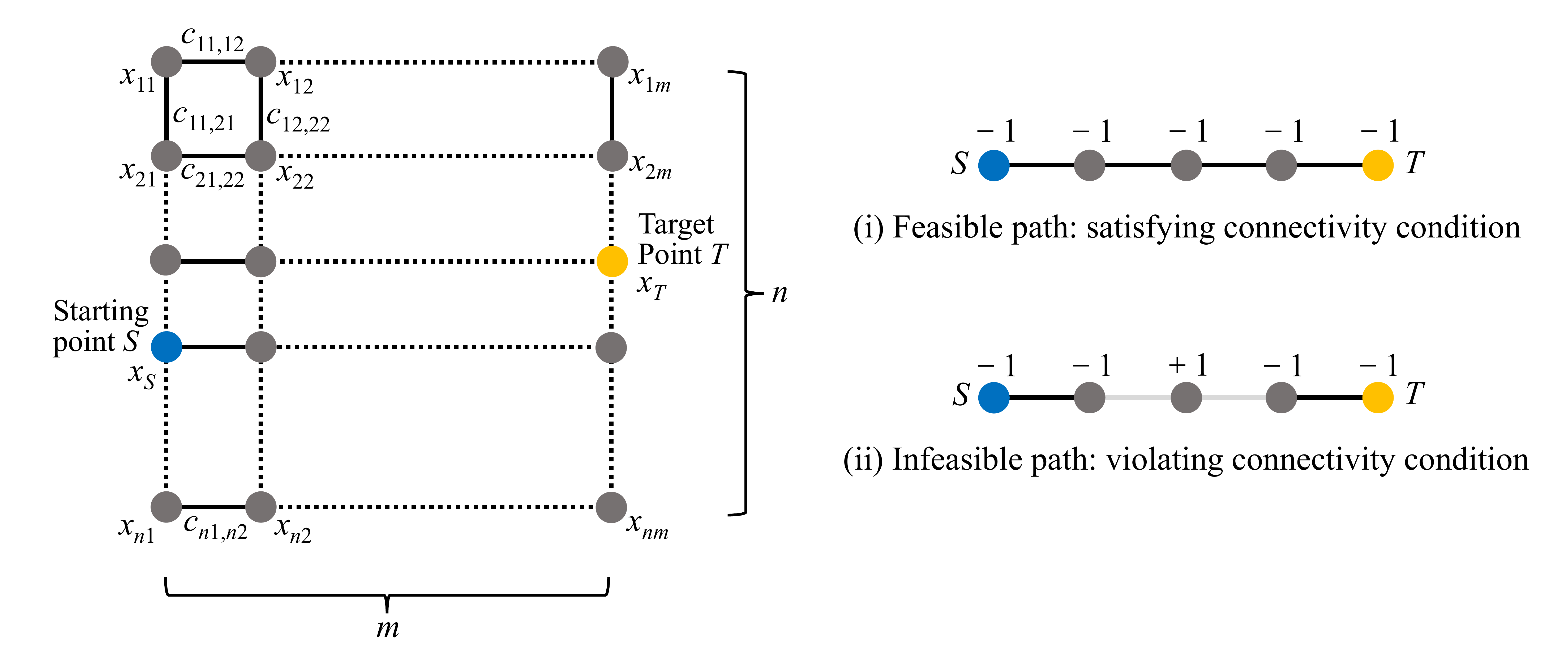}
		\caption{Structural composition of  \(n \times m\) grid map}
		\label{fig:1}
		\end{figure}

	The research object of this study and its structure are illustrated in Figure \ref{fig:1}. There is a two-dimensional grid map with a scale of \(n \times m\) in it. The starting point $S$ is located at any point on the leftmost edge of the grid map, while the target point $T$ is located at any point on the rightmost edge. Any point on the map represents a position that the agent can reach, and the line connecting two points represents a possible movement trajectory of the agent. The number \(c_{i,j}\) (\(c_{i,j} > 0\)) adjacent to the line denotes the path cost from point \(x_i\) to point \(x_j\). When the agent passes through a certain position point, the value of this position point is set to \(-1\); when it does not pass through, the value is set to 1. The values of \(\pm1\) correspond to the two basic spin states of a single qubit in the Ising model, thus laying a foundation for solving this problem using quantum algorithms in subsequent steps.
	
	Under such a setup, if the agent moves from position point \(x_i\) to position point \(x_{i+1}\), then \(x_i = x_{i+1} = -1\), and the corresponding path cost is \(c_{i,i+1}\). Any feasible or infeasible path is defined as a combination of values of all position points in the grid map, with the constraint that both the starting point and the target point must be \(-1\). Among these paths, a feasible path must satisfy the connectivity condition: the agent can only move horizontally or vertically to the next adjacent position point each time, and jumping is not allowed. Based on this, the problem to be solved in this study can be stated as follows: find the optimal feasible path with the minimum total path cost from the starting point $S$ to the target point $T$. The solution process of this problem can be further refined into two specific steps:

	1. Filter and process all path combinations to screen out the set of feasible paths that meet the connectivity condition;

	2. Identify the path with the minimum path cost among all feasible paths, which serves as the solution to the problem.
	
	\section{Characterization and Solution of Connectivity Conditions}
	
\begin{figure}[h]
		\centering
		\includegraphics[width=0.6\linewidth]{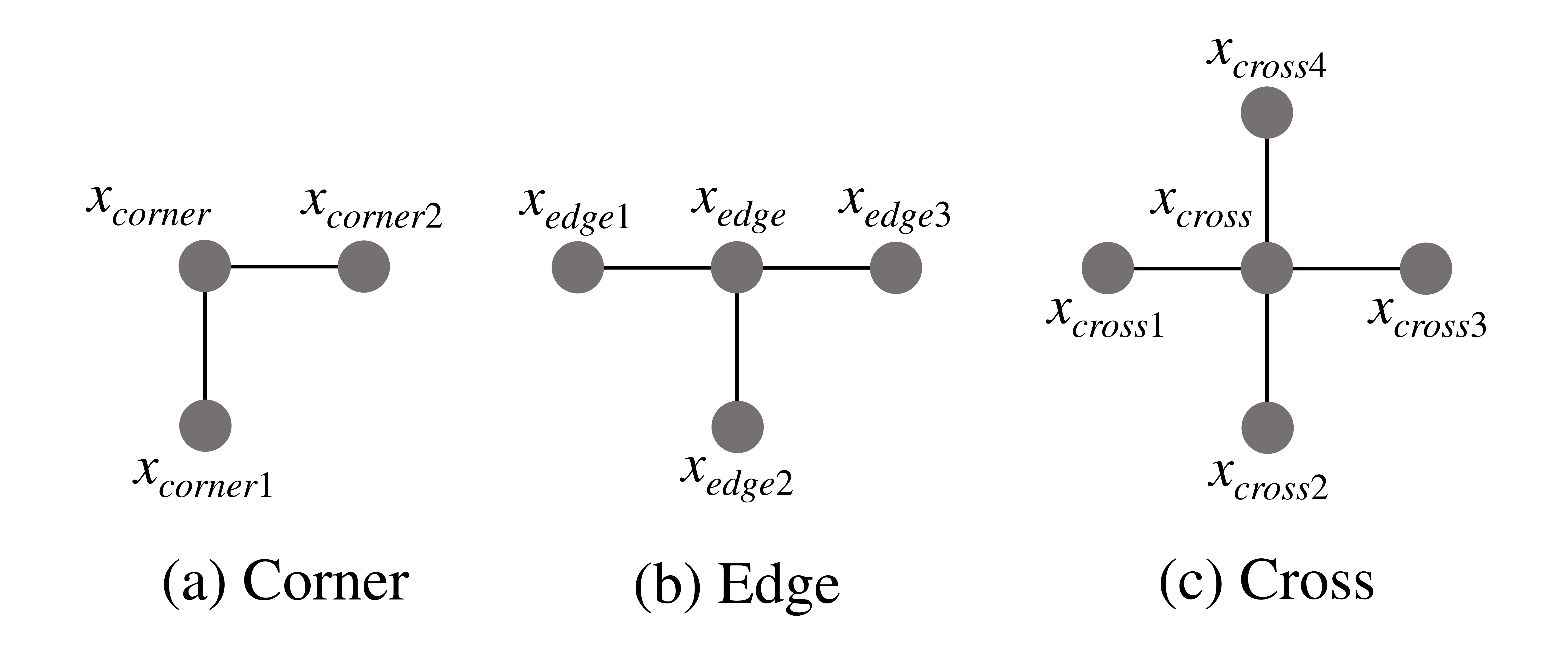}
		\caption{Three fundamental components of \(n \times m\) grid map}
		\label{fig:2}
		\end{figure}

	Inspired by the minimum energy principle, this study transforms the first solution step into a computational process based on the minimum energy principle, which filters out infeasible paths that do not satisfy the connectivity condition. Essentially, all two-dimensional grid maps can ultimately be reduced to three basic structures: `corner', `edge', and `cross', as illustrated in Figure \ref{fig:2}. Among these, `corners' specifically refer to the four outermost corners of the grid map, `edges' denote the outermost edges of the grid map, and `crosses' are enclosed in the middle of the grid map structure. The global connectivity condition for feasible paths in the grid map can be decomposed into the sum of the local connectivity conditions of these three basic structures.
	
	Let \(\mathcal{M}_{corner}\), \(\mathcal{M}_{edge}\), and \(\mathcal{M}_{cross}\) denote the subsets of position nodes composed of the `corner', `edge', and `cross' structures, respectively, after excluding the starting point and target point from the grid map. Let \(\mathcal{M}_S\) and \(\mathcal{M}_T\) represent the subsets of position nodes consisting of the starting point $S$, target point $T$, and their adjacent nodes, respectively. The relationship between these subsets and the total set \(\mathcal{M}\) of position nodes in the grid map is given by:
	\[
	\mathcal{M}_{corner} \cup \mathcal{M}_{edge} \cup \mathcal{M}_{cross} \cup \mathcal{M}_S \cup \mathcal{M}_T = \mathcal{M}
	\]
	
	Further, let \(\mathcal{M}_1\), \(\mathcal{M}_2\), and \(\mathcal{M}_3\) denote the combinations of position nodes randomly selected from the `corner', `edge', and `cross' subsets, respectively, which can form the corresponding basic structures. Mathematically, they are defined as:
	\[
	\mathcal{M}_1 = \{x_{corner}, x_{corner1}, x_{corner2}\}
	\]
	\[
	\mathcal{M}_2 = \{x_{edge}, x_{edge1}, x_{edge2}, x_{edge3}\}
	\]
	\[
	\mathcal{M}_3 = \{x_{cross}, x_{cross1}, x_{cross2}, x_{cross3}, x_{cross4}\}\
	\]
	
	According to the problem setup, the starting point is located on the leftmost edge of the grid, and the target point is on the rightmost edge. Consequently, the starting point, target point, and their adjacent nodes can only form either `corner' or `edge' structures. First, when the subset \(\mathcal{M}_S\) forms a `corner' or `edge' structure, its position node elements are expressed as:
	\[
	\mathcal{M}_S = \{x_S, x_{S1}, x_{S2}\} \quad \text{or} \quad \mathcal{M}_S = \{x_S, x_{S1}, x_{S2}, x_{S3}\}
	\]
	
	Second, when the subset \(\mathcal{M}_T\) forms a `corner' or `edge' structure, its position node elements are given by:
	\[
	\mathcal{M}_T = \{x_T, x_{T1}, x_{T2}\} \quad \text{or} \quad \mathcal{M}_T = \{x_T, x_{T1}, x_{T2}, x_{T3}\}
	\]
	
	The single-element subsets containing only the starting point $S$ and target point $T$ are denoted as \(\mathcal{X}_S = \{x_S\}\) and \(\mathcal{X}_T = \{x_T\}\), respectively. Based on the above partitioning, the connectivity conditions for different basic structures of the grid map are analyzed separately below. The analysis approach involves transforming the process of identifying value combinations of position nodes that satisfy the connectivity condition into a process of solving for the minimum value of the corresponding energy function through mathematical transformations.
	
	\subsection{Connectivity Condition of Corner Paths}
		
		As depicted in Figure \ref{fig:2}, the research object herein is the corner path reference point \(x_{corner}\). When \(x_{corner} = -1\), it signifies that the agent passes through this reference point. In this case, for the corner path to satisfy the connectivity condition, the adjacent position points \(x_{corner1}\) and \(x_{corner2}\) must be activated simultaneously, i.e., \(x_{corner1} = x_{corner2} = -1\). Conversely, when \(x_{corner} = 1\), it indicates that the agent does not pass through this reference point, and the corresponding corner path does not exist—thereby imposing no restrictions on the values of \(x_{corner1}\) and \(x_{corner2}\). Based on this, the connectivity condition for the corner path is constructed in accordance with the minimum energy principle as follows:
		
		\begin{equation}
			\min \ E_1\left(\mathcal{M}_1\right) = \left(1 - x_{corner}\right)\left(x_{corner1} + x_{corner2} + 2\right)
			\label{equation:1}
		\end{equation}
		
		where \(E_1(\mathcal{M}_1)\) represents the energy magnitude of the corner path under different value combinations of position points. When \(x_{corner} = -1\), only when \(x_{corner1} = x_{corner2} = -1\) can Equation \ref{equation:1} attain the minimum energy value, and the connectivity condition for the corner path is satisfied at this moment.
		
		\subsection{Connectivity Condition for Edge Paths}
		
		As shown in Figure \ref{fig:2}, the research object here is the edge path reference point \(x_{edge}\). When \(x_{edge} = -1\), it indicates that the agent passes through this reference point. In this case, for the edge path to satisfy the connectivity condition, the adjacent position points \(x_{edge1}\), \(x_{edge2}\), and \(x_{edge3}\) must be in one of the following two activation scenarios. The first scenario is that only 2 adjacent position points are activated, that is, any 2 of the 3 adjacent position points take the value of $-1$, and the remaining 1 takes the value of 1. The second scenario is that all 3 adjacent position points are activated simultaneously, that is, all take the value of $-1$. When \(x_{edge} = 1\), it means the agent does not pass through this reference point, and the corresponding edge path does not exist—thereby imposing no constraints on the values of \(x_{edge1}\), \(x_{edge2}\), and \(x_{edge3}\) at this time. Therefore, based on the minimum energy principle, the connectivity condition for the edge path is constructed as follows:
			\begin{equation}
				\min \ E_2\left(\mathcal{M}_2\right) = \left(1 - x_{edge}\right)\left[\left(x_{edge1} + x_{edge2} + x_{edge3} + 2\right)^2 - 1\right]
				\label{equation:2}
			\end{equation}
		
		where \(E_2(\mathcal{M}_2)\) characterizes the energy magnitude of the edge path under different value combinations of position points. When \(x_{edge} = -1\), Equation \ref{equation:2} can attain the minimum energy value only if all of \(x_{edge1}\), \(x_{edge2}\), and \(x_{edge3}\) are equal to $-1$ or any two of them are equal to $-1$, and the connectivity condition for the edge path is satisfied at this time.	
		
		\subsection{Connectivity Condition of Cross Paths}
		
		As shown in Figure \ref{fig:2}, the research object here is the cross path reference point \(x_{cross}\). An auxiliary variable \(y_{cross}\) is introduced, and \(y_{cross} = \pm 1\). When \(x_{cross} = -1\), it indicates that the agent passes through this reference point. At this time, if the path is to satisfy the connectivity condition, there are three activation situations for the adjacent position points \(x_{cross1}\), \(x_{cross2}\), \(x_{cross3}\), and \(x_{cross4}\). In the first situation, only 2 adjacent position points are activated, that is, any 2 of the 4 adjacent position points take the value of $-1$, and the remaining 2 take the value of 1. In the second situation, 3 adjacent position points are activated, that is, any 3 of the 4 adjacent position points take the value of $-1$, and the remaining 1 takes the value of 1. In the third situation, all 4 adjacent position points are activated, that is, all take the value of $-1$. When \(x_{cross} = 1\), it indicates that the agent does not pass through this reference point, and the corresponding cross path does not exist. At this time, there is no requirement for the values of \(x_{cross1}\), \(x_{cross2}\), \(x_{cross3}\), and \(x_{cross4}\). Thus, the connectivity condition of the cross path is constructed according to the minimum energy principle as follows:
		\begin{equation}		
			\begin{split}
				\min E_3(\mathcal{M}_3) &= (1 - x_{cross})\left\{(1 - y_{cross})\left[(x_{cross1} + x_{cross2} + x_{cross3} + x_{cross4} + 1)^2 - 1\right] \right. \\
				&\quad \left. + (1 + y_{cross})(x_{cross1} + x_{cross2} + x_{cross3} + x_{cross4} + 4)\right\}
			\end{split}
			\label{equation:3}
		\end{equation}
	
	Among them, \(E_3(\mathcal{M}_3)\) characterizes the energy magnitude of the cross path under different value combinations of position points. When \(x_{cross} = -1\), Equation \ref{equation:3} can only achieve the minimum energy value when \(x_{cross1}\), \(x_{cross2}\), \(x_{cross3}\), and \(x_{cross4}\) are all equal to $-1$, or any 3 of them are equal to 1, or any 2 of them are equal to $-1$. At this point, the connectivity condition of the cross path is satisfied.
	
		\subsection{Connectivity Conditions for Starting and Target points}
	As shown in Figs\ref{fig:1} and\ref{fig:2}, according to the problem setup, the structures of the starting point and the target point can only be `corner' or `edge', and the `cross' structure cannot appear. Moreover, as the starting point, the agent must pass through it, so it must be activated and can only be equal to $-1$. First, consider the starting point \(x_S\). When \(x_S\) belongs to the `corner' structure, there are two activation situations for its 2 surrounding adjacent points, that is, being activated simultaneously or 1 of them being activated. Thus, both can be $-1$ or only 1 of them can be $-1$. When \(x_S\) belongs to the `edge' structure, there are three activation situations for its 3 surrounding adjacent points, that is, 1, 2, or 3 of them are activated, i.e., 1, 2, or 3 of them are $-1$. Introduce auxiliary variables \(y_S = \pm 1\) and \(y_T = \pm 1\). According to the minimum energy principle, the connectivity condition for the starting point path is constructed as follows:
	\begin{equation}
	\min E_4(\mathcal{M}_S) = 
	\begin{cases}
		(x_{S1} + x_{S2} + 1)^2 - 1 & \text{if } \mathcal{M}_S \text{ is corner} \\
		\begin{aligned}
			&(1 - y_S)\left[(x_{S1} + x_{S2} + x_{S3})^2 - 1\right] + \\
			&(1 + y_S)(x_{S1} + x_{S2} + x_{S3} + 3)
		\end{aligned} & \text{if } \mathcal{M}_S \text{ is edge}
	\end{cases}	
	\label{equation:4}
	\end{equation}
		
	The derivation logic for the target point \(x_T\) is the same as that for the starting point \(x_S\). Therefore, according to the minimum energy principle, the corresponding connectivity condition for the target point path is constructed as follows:
	\begin{equation}
		\min E_4(\mathcal{M}_T) = 
		\begin{cases}
			(x_{T1} + x_{T2} + 1)^2 - 1 & \text{if } \mathcal{M}_T \text{ is corner} \\
			\begin{aligned}
				&(1 - y_T)\left[(x_{T1} + x_{T2} + x_{T3})^2 - 1\right] + \\
				&(1 + y_T)(x_{T1} + x_{T2} + x_{T3} + 3)
			\end{aligned} & \text{if } \mathcal{M}_T \text{ is edge}
		\end{cases}
		\label{equation:5}
	\end{equation}	
	
	To ensure that the agent passes through the starting point and the target point, both should simultaneously take the minimum value of $-1$, that is, be activated at the same time. Based on the minimum energy principle, the corresponding connectivity condition is constructed as follows:
	
\begin{equation}
		\min E_6(\mathcal{X}_S, \mathcal{X}_T) = x_S + x_T + 2
		\label{equation:6}
	\end{equation}
		
	After combining the above six energy functions that characterize the connectivity conditions, the total connectivity energy function \(E(\mathcal{M}_{123456})\) can be obtained:
	\begin{equation}
		\begin{split}
			E(\mathcal{M}_{123456}) &= \sum_{\mathcal{M}_1 \subseteq \mathcal{M}_{corner}} E_1(\mathcal{M}_1) + \sum_{\mathcal{M}_2 \subseteq \mathcal{M}_{edge}} E_2(\mathcal{M}_2) \\
			&\quad + \sum_{\mathcal{M}_3 \subseteq \mathcal{M}_{cross}} E_3(\mathcal{M}_3) + E_4(\mathcal{M}_S) + E_5(\mathcal{M}_T) + E_6(\mathcal{X}_S, \mathcal{X}_T)
		\end{split}
		\label{equation:7}
	\end{equation}	
	
	where \(\mathcal{M}_{123456}\) represents the union formed by the value combinations of the position points of the three basic structures, together with the value combinations of the position points of the starting and target structures and the values of the starting and target points themselves:
	\[
	\mathcal{M}_{123456} = \mathcal{M}_1 \cup \mathcal{M}_2 \cup \mathcal{M}_3 \cup \mathcal{M}_S \cup \mathcal{M}_T \cup \mathcal{X}_S \cup \mathcal{X}_T
	\]
	
	Based on the above process, the problem of finding a feasible path that satisfies the connectivity condition in the grid map is transformed into the problem of solving the minimum value of the connectivity energy function \(E(\mathcal{M}_{123456})\). Only those position point combinations that make the connectivity energy function take the minimum value of 0 are feasible paths that satisfy the connectivity condition.
	
	\section{Overall Solution Process}
	Let the set composed of feasible paths be \(\mathcal{F}\). Then, the first step of solving the problem can be further specified as follows: from the set of all feasible and infeasible paths \(\mathcal{M}\), select the combination of position points \(\mathcal{M}^*_{123456}\) that minimizes the connectivity energy function \(E(\mathcal{M}_{123456})\); this combination corresponds to the feasible path, as shown in Equation \ref{equation:8}:
		\begin{equation}
			\begin{aligned}
				&\mathcal{F} \supseteq {path} = \mathcal{M}^*_{123456} = \arg\min_{\mathcal{M}_{123456}} E\left( \mathcal{M}_{123456} \right) \\
				&\text{subject to } \mathcal{M}_{123456} \subseteq \mathcal{M}
			\end{aligned}
			\label{equation:8}
		\end{equation}
		
		On this basis, the second solution step is executed: from the set of all feasible paths \(\mathcal{F}\), find \({path}^*\) that minimizes the path energy function \(E_c({path})\). The corresponding optimization problem can be expressed as Equation \ref{equation:9}:
		\begin{equation}
			\begin{aligned}
				&path^* = \arg\min_{path} E_c\left(path\right) = \arg\min_{path} \left\{ \frac{1}{8} \sum_{x_i, x_{neighbor} \in \mathcal{F}} c_{i,neighbor} \left[ \left(x_i + x_{neighbor} - 1\right)^2 - 1 \right] \right\} \\
				&\text{subject to } path \subseteq \mathcal{F}
			\end{aligned}
			\label{equation:9}
		\end{equation}
		
		In Equation \ref{equation:9}, \(x_{neighbor}\) represents any position point adjacent to the position point \(x_i\), and \(c_{i,{neighbor}}\) denotes the cost incurred when the agent moves from the position point \(x_i\) to the adjacent position point \(x_{neighbor}\). The coefficient \(1/8\) is a normalization coefficient, whose function is to convert the calculation result of the expression into the actual passage cost. When the agent passes through the two position points in Equation \ref{equation:9}, where \(x_i = x_{neighbor} = -1\), the above expression is exactly equal to the passage cost \(c_{i,{neighbor}}\). Conversely, when the agent does not pass through the two position points, the passage cost is 0. The optimal path \({path}^*\) therein is the solution to the problem.
		
		The above optimal path solving problem for grids is essentially an NP-complete problem. Although classical algorithms can be used to solve the optimal feasible path for small-scale grid maps, as the number of position points increases, the computational difficulty of classical methods will grow exponentially. When the number of position points reaches a certain threshold, it will be impossible to solve the problem using existing classical devices. In view of this, this paper attempts to break through the above bottleneck by leveraging the advantages of quantum algorithms and solve the grid path planning problem based on the minimum energy principle.
		
		\section{Solution Based on Parallel QAQA Architecture}
		\subsection{General Solution Framework of Parallel QAQA Architecture}
		Let \(H(t)\) be a time-varying Hamiltonian. Then, the Schrödinger equation is given by:
		\[
		\mathrm{i}\hbar \frac{\partial}{\partial t} \ket{\varphi(t)} = H(t) \ket{\varphi(t)}
		\]
		
		By solving the Schrödinger equation, the quantum state evolution equation can be obtained:
		\[
		\ket{\varphi(t)} = \exp\left[ -\frac{\mathrm{i} H(t)}{\hbar}\right] \ket{\varphi(0)}
		\]
		
		By setting the reduced Planck constant \(\hbar\) to 1, the quantum state evolution equation is simplified as:
		\[
		\ket{\varphi(t)} = \exp[-\mathrm{i}H(t)]\ket{\varphi(0)}
		\]
		
		According to the quantum adiabatic algorithm\cite{Farhi2000}, the Hamiltonian \(H(t)\) in the above equation can be further decomposed into:
		\[
		H(t)=\left(1 - \frac{t}{T}\right)H_0+\frac{t}{T}H_1\
		\]
		
		where \(T\) is the total evolution time, \(H_0\) is the initial Hamiltonian, and \(H_1\) is the Hamiltonian encoding the target problem. According to the solution idea of QAOA, the above evolution process can be expressed by a discretized approximate processing method\cite{Farhi2014}:
		\begin{equation}
			\exp\left(\mathrm{i}\beta_p H_0\right)\exp\left(\mathrm{i}\gamma_p H_1\right)\dots\exp\left(\mathrm{i}\beta_2 H_0\right)\exp\left(\mathrm{i}\gamma_2 H_1\right)\exp\left(\mathrm{i}\beta_1 H_0\right)\exp\left(\mathrm{i}\gamma_1 H_1\right)\ket{\varphi(0)}
			\label{equation:10}
		\end{equation}
		
		In Equation \ref{equation:10}, \(p\) represents the number of discrete evolution subprocesses into which the continuous evolution process is decomposed, that is, the number of layers in the QAOA circuit. The corresponding vectors \(\boldsymbol{\beta} = (\beta_1, \beta_2, \dots, \beta_p)\) and \(\boldsymbol{\gamma} = (\gamma_1, \gamma_2, \dots, \gamma_p)\) denote the sets composed of two groups of coefficient parameters. Assuming that the entire circuit contains \(l\) qubits in total, the initial quantum state is:
		\[
		\ket{\varphi(0)} = \mathrm{H}^{\otimes l} \ket{0}^{\otimes l} = \ket{+}^{\otimes l} = \frac{1}{\sqrt{2^l}} \sum_{i=0}^{2^l - 1} \ket{i}
		\]
		
		In the equation, \(\mathrm{H}\) is the Hadamard gate. The initial Hamiltonian \(H_0\) is denoted as:
		\[
		H_0 = \sum_{k=1}^{l} \sigma_x^k
		\]
		
		Here, \(\sigma_x^k\) denotes the application of the Pauli-X gate to the $k$-th qubit, while the identity gate \(\text{I}\) is applied to the remaining qubits. After constructing the QAOA circuit according to the above solution process, first, the values of vectors \(\boldsymbol{\beta}\) and \(\boldsymbol{\gamma}\) are randomly determined. Then, they are brought into the circuit to perform the approximate quantum evolution process and conduct measurements, thereby solving the energy value corresponding to the target Hamiltonian \(H_1\):
		\begin{equation}
			\langle \boldsymbol{\gamma}, \boldsymbol{\beta} | H_1 | \boldsymbol{\gamma}, \boldsymbol{\beta} \rangle = \sum_{i=0}^{2^l - 1} |\alpha_i|^2 \langle i | H_1 | i \rangle \approx \sum_{i=0}^{2^l - 1} \frac{m_i}{M} \langle i | H_1 | i \rangle
			\label{equation:11}
		\end{equation}
		
		In Equation \ref{equation:11}, $M$ represents the total number of measurements, and \(m_i\) denotes the number of times the quantum state \(|i\rangle\) is measured. Subsequently, a classical algorithm is employed to solve the gradient of the energy value, and the gradient is used to update the values of \(\boldsymbol{\beta}\) and \(\boldsymbol{\gamma}\). The above update process is executed repeatedly until the minimum energy value is obtained. At this point, the quantum state \(|\boldsymbol{\beta}^*, \boldsymbol{\gamma}^*\rangle\) determined by the optimal vectors \(\boldsymbol{\beta}^*\) and \(\boldsymbol{\gamma}^*\) corresponds to the minimum energy value of the target Hamiltonian \(H_1\). Finally, by bringing \(\boldsymbol{\beta}^*\) and \(\boldsymbol{\gamma}^*\) into the QAOA circuit and executing it, the approximately optimal quantum state corresponding to the target problem can be obtained through measurement, which is the approximately optimal solution to the problem.
		
		Transforming the target problem to be solved into the corresponding Hamiltonian \(H_1\) is a key step in constructing the QAOA algorithm circuit. Combining with the problem studied in this paper, first, the connectivity energy function \(E(\mathcal{M}_{123456})\) and the path energy function \(E_c({path})\) are fully expanded and converted into an energy polynomial composed of combinations of position point variables. Next, all position point variables are sequentially transformed into corresponding qubits. Third, the position point variables contained in the energy polynomial are transformed into the tensor product of the Pauli-Z gate \(\sigma_z\) and the identity gate \(\text{I}\), so as to obtain the corresponding Hamiltonian. The specific rule is to transform the position point variables in the polynomial into Pauli-Z gates \(\sigma_z\) item by item according to their serial numbers, and the remaining vacant positions are transformed into identity gate \(\text{I}\). Since the qubit can only take the value \(|0\rangle\) or \(|1\rangle\), the effects of the Pauli-Z gate on them are respectively:
		\[
		\sigma_z |0\rangle = |0\rangle, \sigma_z |1\rangle = -|1\rangle
		\]
		
		Therefore, when the eigenvector is the qubit \(|0\rangle\), the eigenvalue of the Pauli-Z gate \(\sigma_z\) is 1, which corresponds to the situation where the agent in the grid map does not pass through this position point and the position point is not activated. Conversely, when the eigenvector is the qubit \(|1\rangle\), the corresponding eigenvalue of the Pauli-Z gate \(\sigma_z\) is \(-1\), which corresponds to the situation where the agent in the grid map passes through this position point and the position point is activated.
		
		For example, suppose a grid map contains a total of \(3 \times 3 = 9\) position points, and the transformation of a certain term \(x_2x_3x_7\) in its energy polynomial is as follows:
		\[
		x_2 x_3 x_7 \to \text{I} \otimes \sigma_z \otimes \sigma_z \otimes \text{I} \otimes \text{I} \otimes \text{I} \otimes \sigma_z \otimes \text{I} \otimes \text{I} = \sigma_z^2 \sigma_z^3 \sigma_z^7
		\]
		
		Assuming the agent passes through the 1st and 2nd points in the map, the corresponding qubit encoding is:
		\[
		\ket{1}^{\otimes 2} \ket{0}^{\otimes 7} = \ket{110000000}
		\]
		
		According to the property of the Pauli-Z gate, the calculation result of \(\sigma_z^2 \sigma_z^3 \sigma_z^7\) acting on the above quantum state is:
		\[
		\sigma_z^2 \sigma_z^3 \sigma_z^7 \ket{110000000} = -\ket{110000000}
		\]
		
		On this basis, by respectively executing the above calculation steps for the connectivity energy function and the path energy function, the Hamiltonian \(H_{11}\) characterizing the total connectivity energy and the Hamiltonian \(H_{12}\) characterizing the total path energy can be obtained, and both contain all the information of the target function to be solved. According to the solution methods in the theory and model part, as well as the mainstream processing methods, \(H_{11}\) and \(H_{12}\) need to be added together to form a complete Hamiltonian \(H_1\), so that calculation and solution can be carried out in the same QAOA circuit. However, under the existing NISQ conditions, the depth and complexity of the quantum circuit have a direct impact on the calculation accuracy\cite{Preskill2018}. When the number of position points in the grid map is large, the depth and complexity of the quantum circuit corresponding to \(H_1\) will also increase accordingly, thereby affecting the calculation accuracy. This paper holds that the theoretical solution process mentioned earlier can be appropriately improved. By constructing two parallel QAOA circuits and executing the two solution processes of connectivity energy and path energy simultaneously, and then using classical algorithms to merge the calculation results of the two, the approximately optimal solution to the problem can finally be obtained.
		
		First, convert the total of \(n \times m\) position point variables in the grid map into corresponding qubits one by one in sequence. Then, construct the quantum computing circuit QAOA1 corresponding to the connectivity energy function. Its initial Hamiltonian is still \(H_0\), but the final Hamiltonian is \(H_{11}\). The corresponding quantum state evolution expression is as follows:
		\begin{equation}
			\exp\left(\mathrm{i}\beta_{1p_1} H_0\right)\exp\left(\mathrm{i}\gamma_{1p_1} H_{11}\right)\dots\exp\left(\mathrm{i}\beta_{12} H_0\right)\exp\left(\mathrm{i}\gamma_{12} H_{11}\right)\exp\left(\mathrm{i}\beta_{11} H_0\right)\exp\left(\mathrm{i}\gamma_{11} H_{11}\right)\ket{\varphi(0)}
			\label{equation:12}
		\end{equation}
		
		The evolution parameters of the circuit are characterized by vectors \(\boldsymbol{\beta}_1\) and \(\boldsymbol{\gamma}_1\), that is, \(\boldsymbol{\beta}_1 = (\beta_{11}, \beta_{12}, \dots, \beta_{1p_1})\) and \(\boldsymbol{\gamma}_1 = (\gamma_{11}, \gamma_{12}, \dots, \gamma_{1p_1})\). The standard QAOA optimization process is executed, and the evolution parameters are updated through the energy gradient to obtain the optimal parameter vectors \(\boldsymbol{\beta}_1^*\) and \(\boldsymbol{\gamma}_1^*\). By bringing this set of optimal parameter vectors into the circuit QAOA1 and performing measurements, the original vector of the occurrence probabilities of all position point value combinations can be obtained:
		\begin{equation}
			\hat{\boldsymbol{P}}_1 = \left( \hat{P}_1^0, \hat{P}_1^1, \dots, \hat{P}_1^i, \dots, \hat{P}_1^{2^{n \times m} - 1} \right) = \left( \frac{m_1^0}{M_1}, \frac{m_1^1}{M_1}, \dots, \frac{m_1^i}{M_1}, \dots, \frac{m_1^{2^{n \times m} - 1}}{M_1} \right)
			\label{equation:13}
		\end{equation}
		
		In Equation \ref{equation:13}, \(M_1\) is the total number of measurements, and \(m_1^i\) is the number of times the quantum state \(|i\rangle\) is measured. Subsequently, to thoroughly filter out the value combinations of position points with very low occurrence probabilities and at the same time highlight the value combinations with high occurrence probabilities more prominently, a threshold probability \(\theta\) (\(0 \leq \theta \leq 1\)) is set. Classical devices are invoked to construct a filter, and the entries in the original probability vector are updated according to the following rules:
		\begin{equation}
			P_1^i \leftarrow \frac{\frac{1}{2}\left( \hat{P}_1^i - \theta + \left| \hat{P}_1^i - \theta \right| \right)}{\sum_{i=0}^{2^{n \times m} - 1} \frac{1}{2}\left( \hat{P}_1^i - \theta + \left| \hat{P}_1^i - \theta \right| \right)}
			\label{equation:14}
		\end{equation}
		
		Thus, the probability vector with enhanced distinguishability for the original measurement probability vector can be obtained as \(\boldsymbol{P}_1 = \left( P_1^0, P_1^1, \dots, P_1^i, \dots, P_1^{2^{n \times m} - 1} \right)\). In this vector, the value combinations of position points with occurrence probabilities lower than the threshold \(\theta\) will be completely filtered out, and the probability of finding them is 0.
		
		At the same time, a circuit QAOA2 corresponding to the path energy function is constructed. Its initial Hamiltonian is also \(H_0\), and the final Hamiltonian is \(H_{12}\). The corresponding evolution expression is as follows:
		\begin{equation}
			\exp\left(\mathrm{i}\beta_{2p_2} H_0\right)\exp\left(\mathrm{i}\gamma_{2p_2} H_{12}\right)\dots\exp\left(\mathrm{i}\beta_{22} H_0\right)\exp\left(\mathrm{i}\gamma_{22} H_{12}\right)\exp\left(\mathrm{i}\beta_{21} H_0\right)\exp\left(\mathrm{i}\gamma_{21} H_{12}\right)\ket{\varphi(0)}
			\label{equatuin:15}
		\end{equation}
		
		The evolution parameters of the circuit are denoted by vectors \(\boldsymbol{\beta}_2\) and \(\boldsymbol{\gamma}_2\), where \(\boldsymbol{\beta}_2 = (\beta_{21}, \beta_{22}, \dots, \beta_{2p_2})\) and \(\boldsymbol{\gamma}_2 = (\gamma_{21}, \gamma_{22}, \dots, \gamma_{2p_2})\). By implementing the standard QAOA optimization procedure, the evolution parameters are updated using the energy gradient to acquire the optimal parameter vectors \(\boldsymbol{\beta}_2^*\) and \(\boldsymbol{\gamma}_2^*\). Substituting this set of optimal parameter vectors into the circuit QAOA2 and conducting measurements, the original vector of occurrence probabilities for all combinations of position point values can be obtained:
		\[
			\boldsymbol{P}_2 = \left( P_2^0, P_2^1, ..., P_2^i, ..., P_2^{2^{n \times m} - 1} \right) = \left( \frac{m_2^0}{M_2}, \frac{m_2^1}{M_2}, ..., \frac{m_2^i}{M_2}, ..., \frac{m_2^{2^{n \times m} - 1}}{M_2} \right)	
		\]
		
		Among them, \(M_2\) represents the total number of measurements, and \(m_2^i\) denotes the number of times the quantum state \(|i\rangle\) is measured. After filtering out the infeasible paths, finding the optimal feasible path with the minimum passing cost is essentially equivalent to finding the intersection of the set of position points that minimize the connectivity energy and the set of position points that minimize the path energy. The probability of its occurrence satisfies the multiplication principle. Since QAOA1 and QAOA2 are two mutually isolated computational circuits, event A (`connectivity filtering') and event B (`path cost minimization') are independent of each other. The probability of the intersection of the two events occurring is equal to the product of their respective occurrence probabilities:
		\[
		\mathrm{Prob}(\mathrm{A} \cap \mathrm{B}) = \mathrm{Prob}(\mathrm{A}) \mathrm{Prob}(\mathrm{B})
		\]
		
		Therefore, by solving the Hadamard product of vectors \(\boldsymbol{P}_1\) and \(\boldsymbol{P}_2\), the original probability solution vector for the optimal path planning problem can be obtained:
		\begin{equation}
		\boldsymbol{\hat{P}} = \left( \hat{P}_0, \hat{P}_1, ..., \hat{P}_i, ..., \hat{P}_{2^{n \times m} - 1} \right) = \boldsymbol{P}_1 \odot \boldsymbol{P}_2 = \left( P_{1}^{0} \cdot P_{2}^{0}, P_{1}^{1} \cdot P_{2}^{1}, ..., P_{1}^{i} \cdot P_{2}^{i}, ..., P_{1}^{2^{n \times m} - 1} \cdot P_{2}^{2^{n \times m} - 1} \right)
		\label{equation:16}
		\end{equation}
		
		Ultimately, by performing normalization on the original probability solution vector, the approximately optimal solution vector for the path planning problem can be obtained:
		\begin{equation}
			\boldsymbol{P} = \frac{\hat{\boldsymbol{P}}}{\sum_{i=0}^{2^{n \times m} - 1} \hat{P}_i} = \frac{1}{\sum_{i=0}^{2^{n \times m} - 1} \hat{P}_i} \left( \hat{P}_1, \hat{P}_2, ..., \hat{P}_i, ..., \hat{P}_{2^{n \times m} - 1} \right)
			\label{equation:17}
		\end{equation}
		
		In Equation \ref{equation:17}, the component of \(\boldsymbol{P}\) with the maximum probability value corresponds to the quantum state encoding of the approximately optimal feasible path from the starting point to the target point, which is the approximately optimal solution to the two-dimensional grid path planning problem.
		
		\subsection{Solving Example for \(2 \times 3\) Grid Map}
		
\begin{figure}[h]
			\centering
			\includegraphics[width=0.4\linewidth]{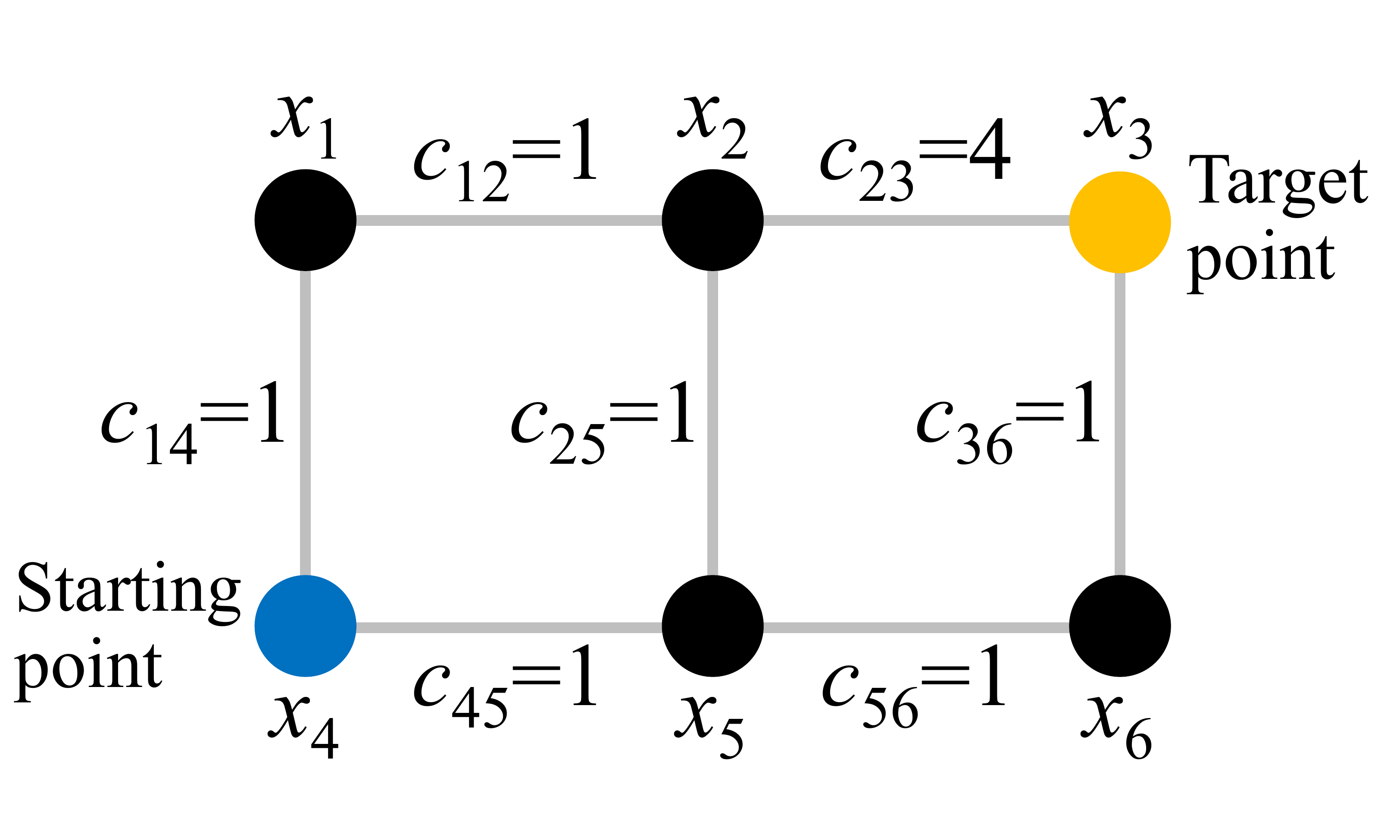}
			\caption{Structure setup of \(2 \times 3\) grid map}
			\label{fig:3}
		\end{figure}

		Taking a \(2 \times 3\) grid map as an example, its structure setup is illustrated in Figure \ref{fig:3}. 
		
\begin{figure}[h]
			\centering
			\includegraphics[width=1\linewidth]{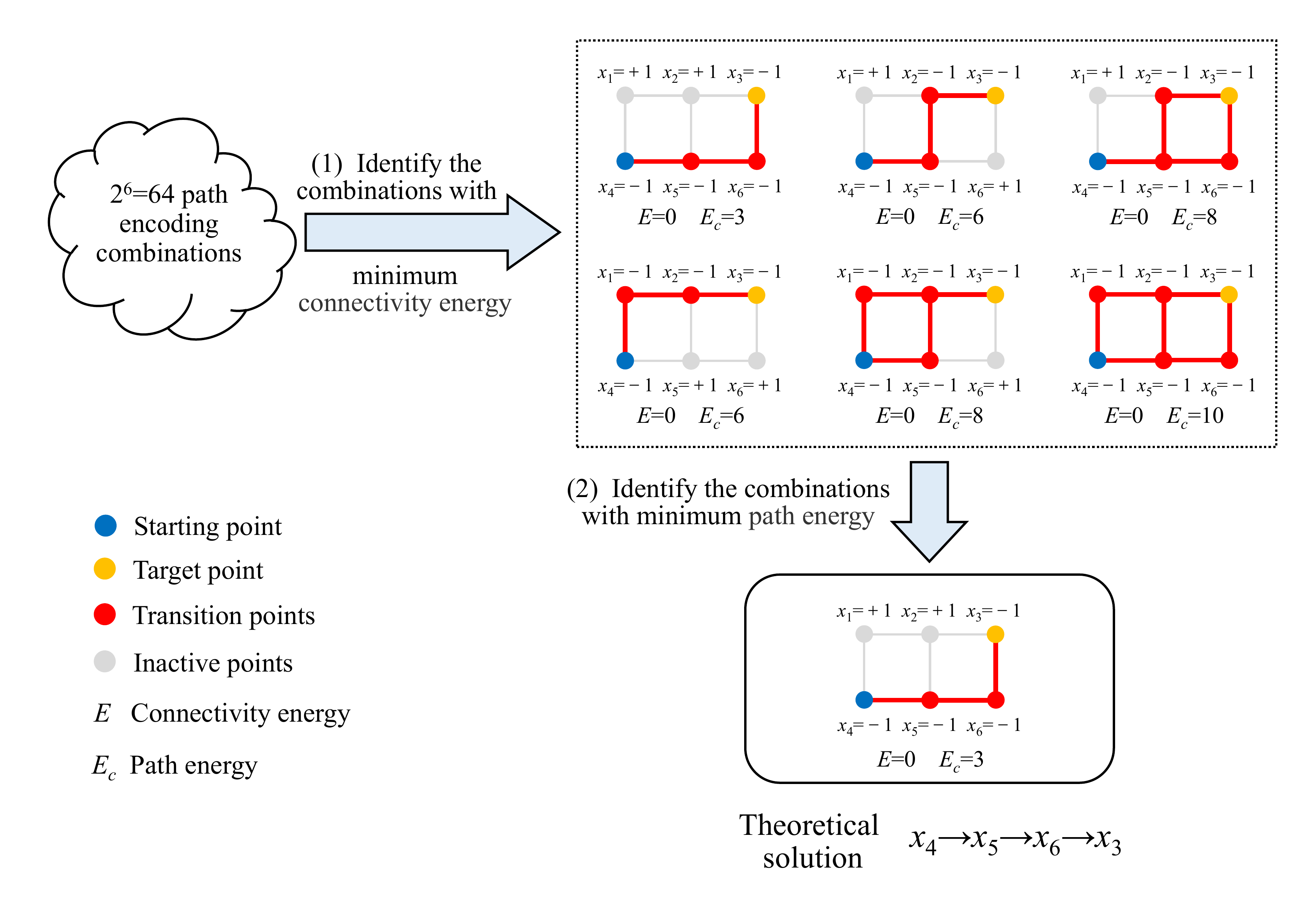}
			\caption{Solution steps for the optimal feasible path of \(2 \times 3\) grid map based on classical algorithms}
			\label{fig:4}
		\end{figure}

		Figure \ref{fig:4} intuitively demonstrates the detailed steps of solving the optimal path for a \(2 \times 3\) grid map using classical algorithms. Starting from the principle of energy minimization, all feasible paths satisfying the connectivity condition are first identified, and then the optimal feasible path with the lowest passing cost is identified. According to the calculation results in Figure \ref{fig:4}, the value conditions of the position points corresponding to the optimal feasible path are: \(x_1 = +1\), \(x_2 = +1\), \(x_3 = -1\), \(x_4 = -1\), \(x_5 = -1\), \(x_6 = -1\). At this time, the order in which the agent moves from the starting point to the target point is: \(x_4 \to x_5 \to x_6 \to x_3\).
		
		The solution process of the parallel QAOA circuits for the optimal path based on the energy minimization principle is explained as follows. According to the analysis in the theory and model section, Figure \ref{fig:3} identifies all position points and their connectivity costs in the grid map. There are a total of 6 position points and 7 connectivity cost variables, and 2 parallel QAOA circuits need to be constructed for calculation processing, where each circuit contains 6 qubits. The corresponding initial Hamiltonian \(H_0\) is denoted as:
		\[
		H_0 = \sum_{k=1}^{6} \sigma_x^k
		\]
		
		On this basis, following the solution scheme of the general solution framework, first list the Hamiltonian \(H_{11}\) corresponding to the connectivity energy, and then list the Hamiltonian \(H_{12}\) corresponding to the total path energy. First, solve the Hamiltonian corresponding to the connectivity energy. There is no `cross' structure in the \(2 \times 3\) grid map, only `corner' and `edge' structures exist. Therefore, the connectivity energy function described by Equation \ref{equation:7} is simplified as:
		\[
		E\left( \mathcal{M}_{123456} \right) = \sum_{\mathcal{M}_1 \subseteq \mathcal{M}_{corner}} E_1\left( \mathcal{M}_1 \right) + \sum_{\mathcal{M}_2 \subseteq \mathcal{M}_{edge}} E_2\left( \mathcal{M}_2 \right) + E_4\left( \mathcal{M}_{S} \right) + E_5\left( \mathcal{M}_{T} \right) + E_6\left( \mathcal{X}_{S}, \mathcal{X}_{T} \right)
		\]
		
		As illustrated in Figure \ref{fig:3}, among the six position point variables, there are 2 `corner' position variables, namely \(x_1\) and \(x_6\), and 2 `edge' position variables, namely \(x_2\) and \(x_5\). The starting point variable is \(x_4\) and the target point variable is \(x_3\).
		
		By observing the `corner' position variables \(x_1\) and \(x_6\) in Figure \ref{fig:3}, it can be found that among their two adjacent points, each contains one starting point or target point. Since the starting point variable and the target point variable must be \(-1\), only the value of the remaining one adjacent point can be changed. Thus, according to Equation \ref{equation:1}, the connectivity energy function describing the corner position variables at this time is:	
		\[
		\begin{aligned}
			\sum_{\mathcal{M}_1 \subseteq \mathcal{M}_{\text{corner}}} E_1\left( \mathcal{M}_1 \right) &= \left( 1 - x_1 \right)\left( x_2 + 1 \right) + \left( 1 - x_6 \right)\left( x_5 + 1 \right) \\
			&= -x_1 + x_2 + x_5 - x_6 - x_1 x_2 - x_5 x_6 + 2
		\end{aligned}
		\]
		
		Among the three adjacent points of the `edge' position variables \(x_2\) and \(x_5\) in Figure \ref{fig:3}, each contains one starting point or target point. Thus, \(x_2\) and \(x_5\) themselves are also adjacent points of the starting point or target point, and are governed by the value rules of the adjacent points of the starting point or target point. In addition, the remaining two adjacent points of \(x_2\) and \(x_5\) are both adjacent points of `corner' position points and adjacent points of the starting point and target point, and are governed by the corresponding value rules. Therefore, under the condition of the \(2 \times 3\) map, the values of the `edge' position and its adjacent point variables are completely governed by the value rules of the surrounding position points, and there is no need to separately list the connectivity energy function \(E_2(\mathcal{M}_2)\) of the edge position points.
		
		According to Equation \ref{equation:4} and Equation \ref{equation:5}, by simplifying the expanded form using \(x_i^2 = 1\), the connectivity energy function of the starting point and target point is:
		\[
		\begin{aligned}
			E_4\left( \mathcal{M}_{\text{s}} \right) &= \left( x_1 + x_5 + 1 \right)^2 - 1 \\
			&= 2x_1 + 2x_5 + 2x_1 x_5 + 2
		\end{aligned}
		\]
		
		\[
		\begin{aligned}
			E_5\left( \mathcal{M}_{\text{T}} \right) &= \left( x_2 + x_6 + 1 \right)^2 - 1 \\
			&= 2x_2 + 2x_6 + 2x_2 x_6 + 2
		\end{aligned}
		\]
		
		According to Equation \ref{equation:6}, the energy function describing the specific positions of the starting point and the target point is:
		\[
		\begin{aligned}
			E_{6}\left(\mathcal{X}_{\mathrm{S}}, \mathcal{X}_{\mathrm{T}}\right)=x_{4}+x_{3}+2
		\end{aligned}
		\]
		
		Thus, the total connectivity energy function is:
		\[
		\begin{aligned}
			E\left( \mathcal{M}_{123456} \right) &= \sum_{\mathcal{M}_1 \subseteq \mathcal{M}_{\text{corner}}} E_1\left( \mathcal{M}_1 \right) + E_4\left( \mathcal{M}_{\text{S}} \right) + E_5\left( \mathcal{M}_{\text{T}} \right) + E_6\left( \mathcal{X}_{\text{S}}, \mathcal{X}_{\text{T}} \right) \\
			&= x_1 + 3x_2 + x_3 + x_4 + 3x_5 + x_6 - x_1 x_2 + 2x_1 x_5 + 2x_2 x_6 - x_5 x_6 + 8
		\end{aligned}
		\]

		Accordingly, the Hamiltonian \(H_{11}\) corresponding to the connectivity energy function is:
		\[
		\begin{aligned}
			H_{11} &= \lambda_1 \left( \sigma_z^1 + 3\sigma_z^2 + \sigma_z^3 + \sigma_z^4 + 3\sigma_z^5 + \sigma_z^6 - \sigma_z^1 \sigma_z^2 + 2\sigma_z^1 \sigma_z^5 + 2\sigma_z^2 \sigma_z^6 - \sigma_z^5 \sigma_z^6 \right)
		\end{aligned}
		\]
		
		where \(\lambda_1\) is a multiplication factor, whose role is to enlarge the energy difference under different position point value combinations, so as to effectively identify the combination corresponding to the lowest energy state. In the subsequent computational experiments of this paper, \(\lambda_1\) is set to 100. Meanwhile, to simplify the calculation process, the constant term 8 is omitted. Next, solve the Hamiltonian corresponding to the total path energy. In this paper, the cost \(c_{23}\) for the agent to move from position point 2 to position point 3 is set to 4, and the movement costs between the remaining position points are all set to 1. According to Equation \ref{equation:9}, the path energy function is:
		\[
		\begin{aligned}
			E_c\left( path \right) &= \frac{1}{8} \Big\{ \left[ \left( x_1 + x_2 - 1 \right)^2 - 1 \right] + \left[ \left( x_1 + x_4 - 1 \right)^2 - 1 \right] \\
			&\quad + \left[ \left( x_2 + x_5 - 1 \right)^2 - 1 \right] + 4\left[ \left( x_2 + x_3 - 1 \right)^2 - 1 \right] \\
			&\quad + \left[ \left( x_3 + x_6 - 1 \right)^2 - 1 \right] + \left[ \left( x_4 + x_5 - 1 \right)^2 - 1 \right] \\
			&\quad + \left[ \left( x_5 + x_6 - 1 \right)^2 - 1 \right] \Big\} \\
			&= \frac{1}{8} \left( -4x_1 - 12x_2 - 10x_3 - 4x_4 - 6x_5 - 4x_6 \right. \\
			&\quad + 2x_1 x_2 + 2x_1 x_4 + 8x_2 x_3 + 2x_2 x_5 + 2x_3 x_6 \\
			&\quad \left. + 2x_4 x_5 + 2x_5 x_6 + 20 \right)
		\end{aligned}
		\]
		
		Therefore, the Hamiltonian \(H_{12}\) corresponding to the path energy function is:
		\[
		\begin{aligned}
			H_{12} &= \lambda_2 \left( -2\sigma_z^1 - 6\sigma_z^2 - 5\sigma_z^3 - 2\sigma_z^4 - 3\sigma_z^5 - 2\sigma_z^6 \right. \\
			&\quad + \sigma_z^1 \sigma_z^2 + \sigma_z^1 \sigma_z^4 + 4\sigma_z^2 \sigma_z^3 + \sigma_z^2 \sigma_z^5 + \sigma_z^3 \sigma_z^6 \\
			&\quad \left. + \sigma_z^4 \sigma_z^5 + \sigma_z^5 \sigma_z^6 \right)
		\end{aligned}
		\]
		
		In the equation, \(\lambda_2\) is a multiplication factor. In subsequent computational experiments, \(\lambda_2\) is set to 100. Meanwhile, to simplify the calculation process, the common factor of the polynomial is eliminated, and the constant term 20 is omitted.
		
\begin{figure}
			\centering
			\includegraphics[angle=90,width=0.76\linewidth]{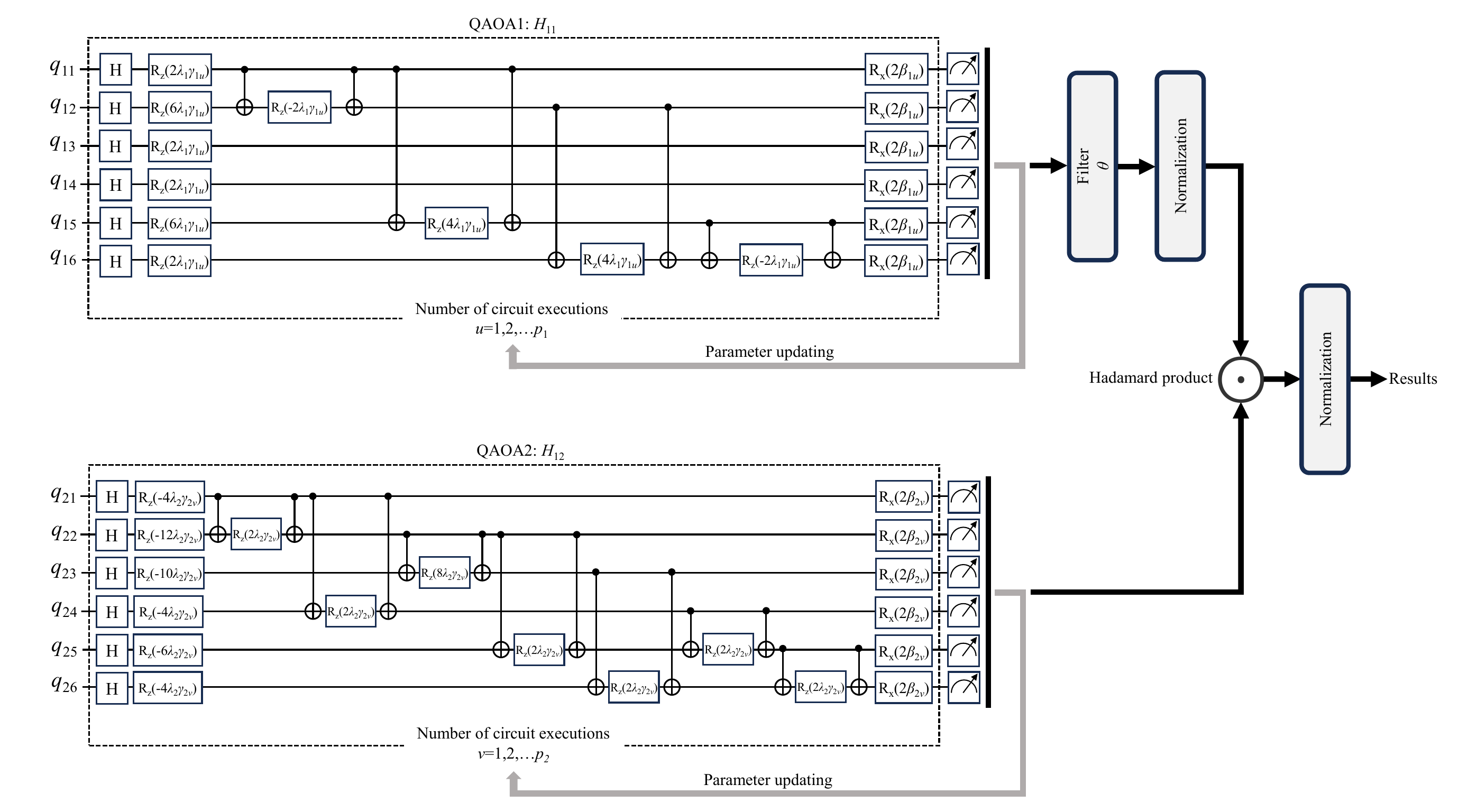}
			\caption{Hybrid circuits for solving the optimal feasible path in a \(2 \times 3\) grid map}
			\label{fig:5}
		\end{figure}

		The second step is to construct a `classical+quantum' hybrid circuit according to the previous calculation steps to solve the optimal path planning problem, as shown in Figure \ref{fig:5}. Among them, \(H_{11}\) is encoded into the QAOA1 circuit, and \(H_{22}\) is encoded into the QAOA2 circuit. A total of \(2 \times 6 = 12\) qubits are invoked in the entire calculation circuit, and the initial state of all qubits is set to \(|0\rangle\).
		
		After executing the computational circuit in Figure \ref{fig:5} and obtaining the final calculation results, the approximate solution of the optimal feasible path can be determined. According to the settings in the theoretical model construction section, the quantum state corresponding to the optimal feasible path at this point is \(|001111\rangle\). This indicates that the agent moves from the starting point \(x_4\) to the target point \(x_3\) following the sequence \(x_4 \to x_5 \to x_6 \to x_3\).
		
		\section{Experimental Results}
		From the theoretical analysis presented earlier, it is known that there are two key variables influencing the final calculation results: the filter parameter \(\theta\) and the number of layers $p$ in the QAOA circuit. To ensure the consistency of the two parallel QAOA circuits, this study sets the number of layers for QAOA1 and QAOA2 to be the same, i.e., \(p_1 = p_2 = p\). Subsequent computational experiments were conducted using the PennyLane quantum simulation platform, focusing on the changes in the above three parameters and examining the impact of each parameter variation on the calculation results.
		
		\subsection{Impact of Filter Parameter \(\theta\) on Calculation Results}
		Theoretically, when \(\theta=0\), the filter has no effect; a larger \(\theta\) value corresponds to a better filtering performance. However, an excessively large \(\theta\) value implies overly strong filtering, which may lead to the elimination of experimental data that meets the requirements. To investigate the impact of different \(\theta\) values on the calculation results in detail, this study designed corresponding computational experiments, where \(\theta\) was set to 6 distinct values starting from 0. To focus specifically on analyzing the effect of \(\theta\), the number of circuit layers $p$ and the number of measurements $shots$ were set as control variables with the following specific values:
		\[
		p_1 = p_2 = p = 10\text{, } shots = 2 \times 10^5\
		\]
		
\begin{figure}[h]
			\centering
			\includegraphics[width=1\linewidth]{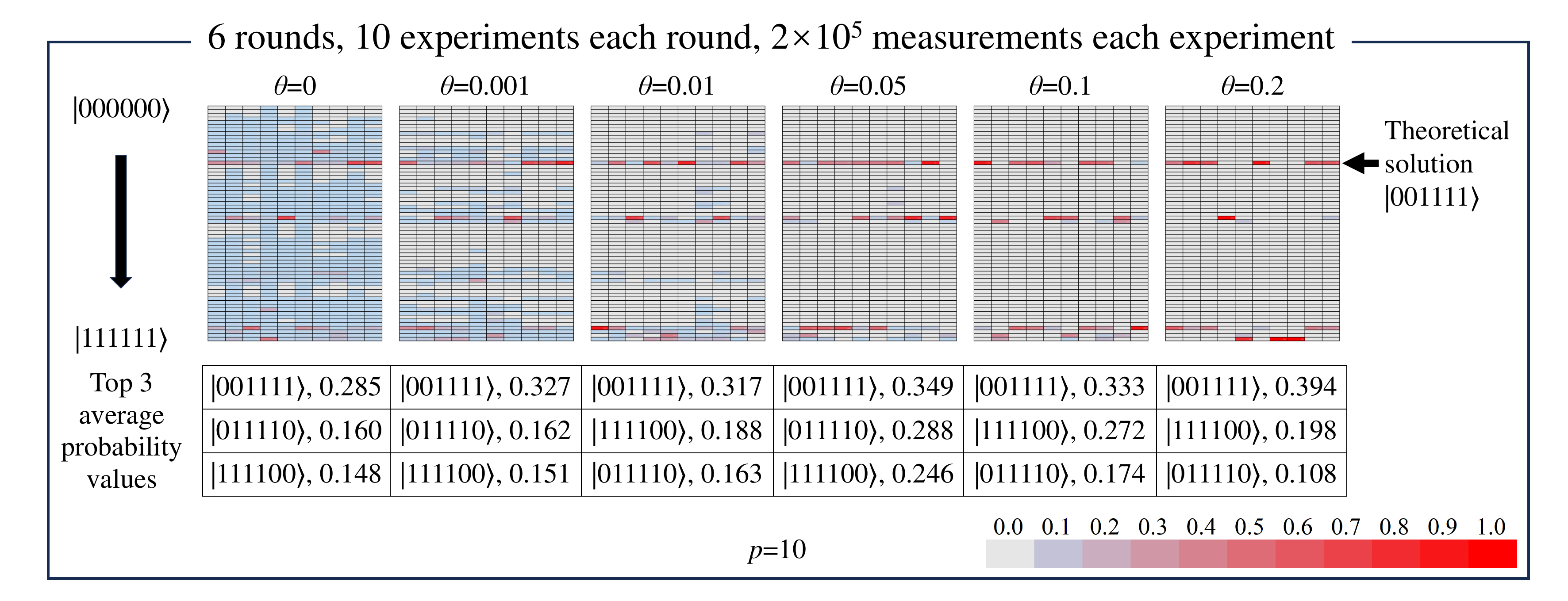}
			\caption{Impact of filter parameter \(\theta\) on calculation results}
			\label{fig:6}
		\end{figure}

		Under each parameter combination, 10 rounds of experiments were conducted, and the experimental results are shown in Figure \ref{fig:6}.
		
		From Figure \ref{fig:6}, it can be observed that when \(\theta=0\), as predicted by the theoretical model, the filter has no effect. At this point, the quantum state \(|001111\rangle\) corresponding to the optimal path is almost submerged in various types of noise. As the value of \(\theta\) gradually increases, the filtering effect begins to manifest. However, under the current NISQ conditions, the original calculation results may inherently contain certain errors. If the filtering effect is excessively strong, it may lead to the filtering out of correct calculation results. Experiments reveal that when \(\theta\) exceeds a specific threshold, the number of measured occurrences of \(|001111\rangle\) decreases significantly. Additionally, experiments further show that when \(\theta > 0.2\), the overly strong filtering effect results in extremely sparse calculation results, making it even difficult to measure valid quantum state combinations. It is thus evident that the filter parameter \(\theta\) must be reasonably selected in practical applications. Based on a comprehensive analysis of the experimental results, the empirical value of \(\theta\) is appropriately set to 0.05.
		
		\subsection{Impact of Circuit Layer Number \textit{p} on Experimental Results}
\begin{figure}[h]
			\centering
			\includegraphics[width=1\linewidth]{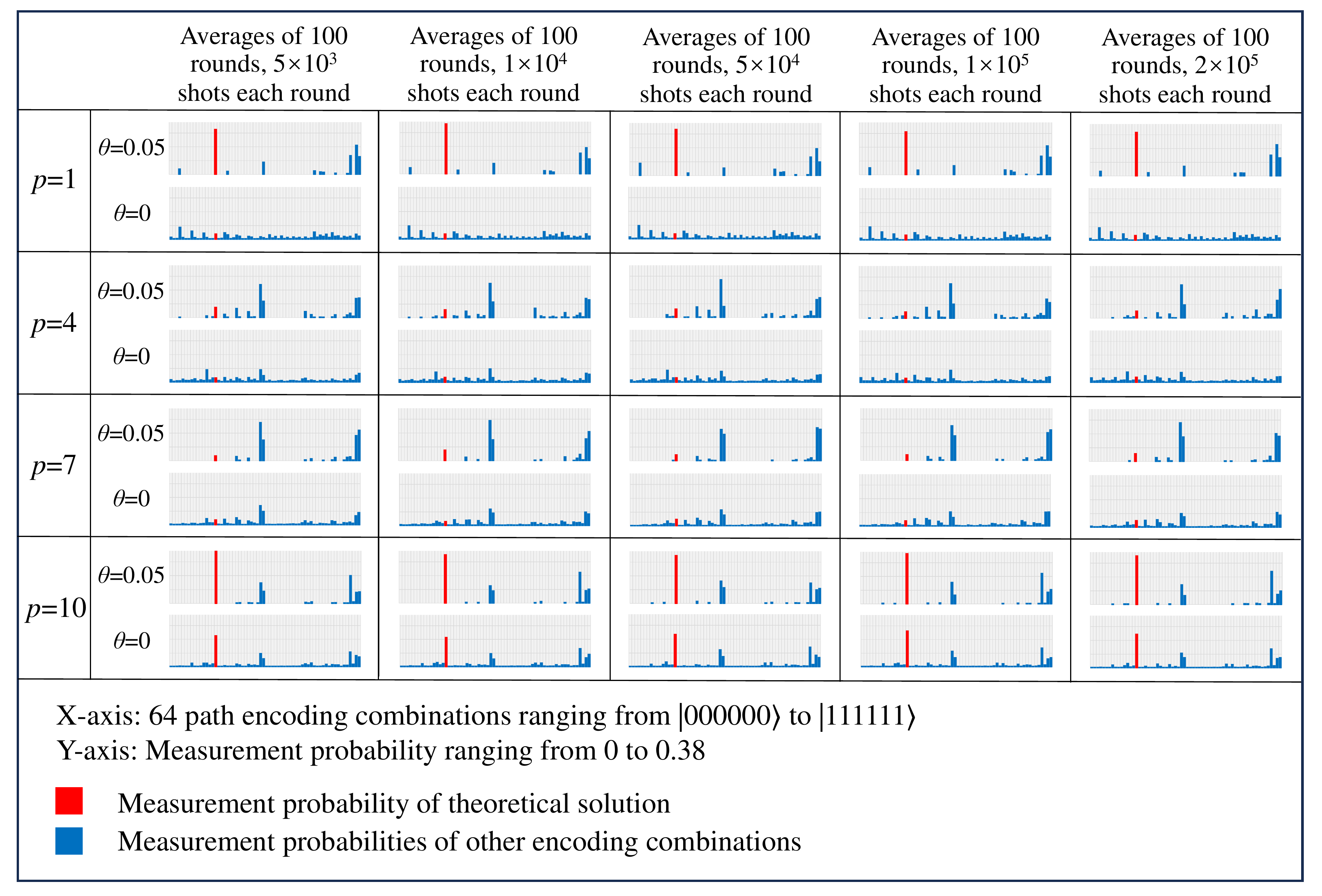}
			\caption{Impact of QAOA circuit layer number $p$ on experimental results}
			\label{fig:7}
		\end{figure}

		To investigate the impact of different circuit layer number $p$ on the calculation results, this study designed computational experiments with 4 distinct sets of $p$ values. Meanwhile, to explore the influence of different measurement counts on the experimental results, 5 groups of incrementally increasing measurement counts were set under each $p$ value; the computational experimental results are presented in Figure \ref{fig:7}. Within each combination box formed by the intersection of a specific $p$ value and measurement count, there are both an experimental group with \(\theta = 0.05\) and a control group with \(\theta = 0\). The histogram in each box represents the average probability of finding each position encoding combination after executing 100 rounds of measurements with the specific shots. Among them, the red bars indicate the probability of measuring \(|001111\rangle\)—the theoretical solution of the optimal feasible path encoding combination—while the blue bars represent the probability of measuring other position encoding combinations. Based on the above settings, a total of \(4 \times 1000 \times (5 \times 10^3 + 1 \times 10^4 + 5 \times 10^4 + 1 \times 10^5 + 2 \times 10^5) = 1.46 \times 10^{10}\) simulated quantum measurements were performed in this study.
		
		The experimental results in Figure \ref{fig:7} indicate the following: (1) When \(p = 1\), by activating the filter, the theoretical solution of the optimal feasible path encoding combination \(|001111\rangle\) can be measured with the highest probability, and the measurement results are highly robust. However, when the filter is inactive, the theoretical solution is submerged in noise. It is thus evident that when the circuit layer number is 1, the theoretical solution of the optimal feasible path encoding combination can be effectively identified by leveraging the critical role of the filter. (2) When \(p = 4\) and \(p = 7\), the theoretical solution of the optimal feasible path encoding combination cannot be measured with the highest probability, regardless of whether the filter is activated. (3) When $p$ further increases to 10, the theoretical solution of the optimal feasible path encoding combination can be measured with the highest probability regardless of the filter’s activation status; however, activating the filter can effectively enhance the probability of measuring this theoretical solution.
		
		The above computational experimental results demonstrate that even when the circuit layer number $p$ is only 1, the theoretical solution of the optimal path encoding combination can be found by fully utilizing the filter. Although the optimal solution can also be measured with the highest probability when \(p = 10\), \(p = 1\) is undoubtedly more practically valuable in terms of saving computing resources and reducing circuit complexity.
		
		\subsection{Control Experiment with Serial Circuits}
		
\begin{figure}[h]
			\centering
			\includegraphics[width=1\linewidth]{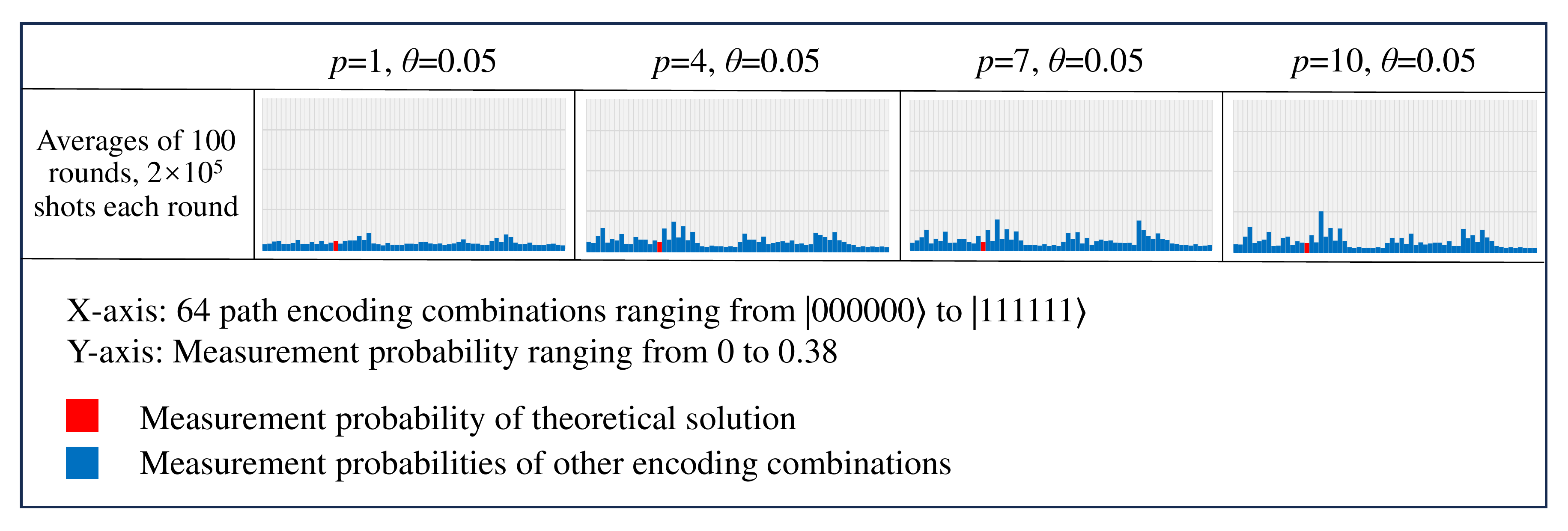}
			\caption{Results of the Serial Circuit Control Experiment}
			\label{fig:8}
		\end{figure}

		To investigate whether there are significant differences between the calculation results of parallel circuits and those of serial circuits, this study conducted a control experiment using serial circuits. The specific experimental design involves converting the parallel QAOA1 and QAOA2 circuits into a serial circuit: on the same row of 6 qubits, QAOA1 is executed first, followed by QAOA2. Figure \ref{fig:8} presents the calculation results that correspond exactly to the respective positions in Figure \ref{fig:7}, under the conditions of a measurement shots of \(2 \times 10^5\), \(\theta = 0.05\), and 4 distinct values of $p$.
		
		By examining the experimental results in Figure \ref{fig:8}, it can be observed that under the serial circuit structure, the theoretical solution of the optimal feasible path encoding combination cannot be identified with the highest measurement probability, and this experimental result is highly robust. This indicates that compared with serial circuits, parallel circuits exhibit a significant advantage: they can find the theoretical solution of the optimal feasible path encoding combination with the highest probability.
		
		\section{Conclusions}
		This study attempts to construct a quantum optimal path solving framework based on parallel QAOA circuits. On the one hand, this framework aims to break the bottleneck of classical schemes in solving NP problems; on the other hand, it seeks to address the challenges faced by mainstream quantum solving frameworks under NISQ conditions.
		The key findings of this research are as follows: (1) By setting an appropriate filter parameter, quantum states of position points with extremely low occurrence probabilities can be effectively filtered out, thereby increasing the probability of obtaining the target quantum state. Based on experimental data, the empirical value of the filter parameter is determined to be 0.05 in this study. (2) When the layer number of the QAOA circuit is only 1, the theoretical solution of the optimal path encoding combination can be identified by fully leveraging the function of the filter. (3) Compared with serial circuits, parallel circuits exhibit a significant advantage: they can find the solution of the optimal feasible path encoding combination with the highest probability.

	\bibliographystyle{unsrtnat} 
	\bibliography{references}
	
\end{document}